\documentclass[epj]{svjour}
\usepackage{graphics}
\usepackage{graphicx}
\usepackage{xspace}
\usepackage{amsmath}

\newcommand{\sqrts}{$\sqrt{s_{_{NN}}}$\xspace}

\newcommand{\pp}{{\it p-p}\xspace}
\begin{document}
\title{Is soft physics entropy driven?}
\author{Helen Caines\inst{1}
}                     
\institute{Yale University, Physics Department, New Haven, CT
 06520}
\date{Received: date / Revised version: date}
%
\abstract{ The soft physics, p$_{T}$ $<$  2 GeV/c, observables at
 both RHIC and the SPS have now been mapped out in
 quite specific detail. From these results there is
 mounting evidence that this regime is primarily driven
 by the multiplicity per unit rapidity, dN$_{ch}$/d$\eta$. This suggests that
 the entropy of the system alone is the underlying driving force
 for many of the global observables measured in heavy-ion collisions.
 That this is the case and there is an apparent independence
 on collision energy is surprising.  I  present the evidence for this multiplicity scaling
 and use it to make some extremely naive predictions for the soft sector
 results at the LHC.
 \PACS{
      {PACS-key}{discribing text of that key}   \and
      {PACS-key}{discribing text of that key}
     } 
} 
\maketitle
\section{Introduction}

\begin{figure}[htbp]
    \begin{center}
  \includegraphics[width=\linewidth]{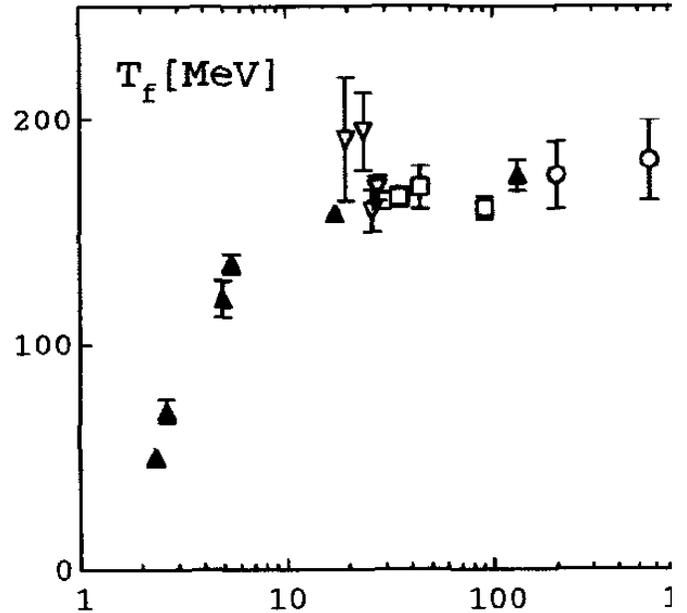}\\
  \caption{The calculated T$_{ch}$ as a function of collision
energy, for e$^{+}$-e$^{-}$ (squares), \pp (open triangles)
$\bar{p}${\it-p} (circles) and A-A (closed triangles). Data from
\cite{TchSqrtS}.}\label{Fig:TChem}
  \end{center}
\end{figure}

Back in 1965, Hagedorn postulated that the hadronic mass spectrum
grows exponentially with mass~\cite{Hagedorn}. Since 1965, more than
3200 additional resonance states have been identified and these results
seem to confirm Hagedorn's hypothesis. The exponential growth means that
adding more  energy to a system merely produces more particles,
resulting in a limiting temperature, T${_0} \sim 160$ MeV for a
hadron gas. Interestingly this value of T$_{0}$ is very close to the
limiting chemical freeze-out temperature, T$_{ch}$, calculated via
statistical production models, from data as a function of \sqrts,
(see Fig.~\ref{Fig:TChem}~\cite{TchSqrtS}). We know that entropy is related to temperature and energy via
T$\Delta$S=$\Delta$E. If the calculations are correct and  chemical
freeze-out temperatures reach a maximal value then, as the collision energy increases, the
entropy of the system must also increase.

~

Even before Hagedorn's work Landau and Fermi related pion production in high energy collisions to
the entropy produced~\cite{Landau,Fermi}. I will try and reproduce the main points of
their arguments here. The energy density, $\epsilon$, available for particle creation in a collision is

\begin{equation}
    \epsilon = \frac{E}{V} = \frac{(\sqrt{s_{NN}} - 2m_{N})\sqrt{s_{NN}}}{2m_{N}V_{0}}
\end{equation}
where: V$_{0}$ is the volume and m$_{N}$ is the mass of the nucleon. If you assume that the entropy is
produced early in the collision, that the source is thermalized and that it expands adiabatically,
the entropy, S, can be related to $\epsilon$ via the equation of state (EoS). Thus, entropy and the
produced particle multiplicity are correlated.

~

A simple example of such a correlation can be derived using the EoS of a system of massless pions which
emits as a blackbody.
In this case, p = $\epsilon$/3 and  $\epsilon$ = T$^{4}$.
Thus
\begin{eqnarray}
        s = S/V \\
        Ts \sim \epsilon + p \\
        s \sim \epsilon ^{3/4}
\end{eqnarray}

and

\begin{eqnarray}
        S \sim V((\sqrt{s_{NN}} - 2m_{N}) \sqrt{s_{NN}})^{3/4} \\
           \sim \frac{N_{part}(\sqrt{s_{NN}} - 2m_{N})^{3/4}}{ \sqrt{s_{NN}}^{1/4}}
           \label{FermiEnergy}
\end{eqnarray}

\begin{figure}[htbp]
 \begin{center}
  \includegraphics[width=\linewidth]{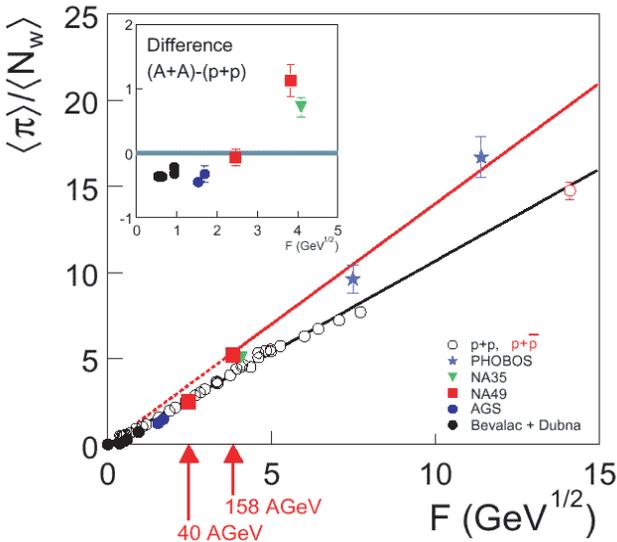}\\
  \caption{Entropy per participant versus the Fermi energy
variable (only pion multiplicities were used) for various collision energies~\cite{EntropyPerPart}.}\label{Fig:Entropy}
   \end{center}
\end{figure}

The \sqrts dependence of the entropy in Eqn.~\ref{FermiEnergy} is the same as that derived by
Fermi~\cite{Fermi} and S/V is called the Fermi energy, F. If these arguments hold, a plot of the
measured entropy density, or mean multiplicity
per participant nucleon, versus F, calculated from the right-hand side of Eqn.~\ref{FermiEnergy} should result in a
linear correlation. Such a correlation can be
seen in Fig.~\ref{Fig:Entropy}, taken from \cite{EntropyPerPart}. In this plot, although only the mean
pion multiplicities are used, this is a good approximation to the total multiplicity. It can be
seen that the multiplicity per participant of the \pp collision data, open circles, is indeed proportional to F. The A-A
data seem to depart from this curve at the top AGS energy. This apparent increase in entropy
in higher energy collisions was first observed by 1995 and interpreted as
evidence of a transition to quark degrees of freedom resulting in a different EoS~\cite{Marek} .

~

Since there appears to be evidence that particle multiplicity and entropy are correlated over a
range of collision energies, it is interesting to
see if entropy, e.g. multiplicity, can be used to determine unique scalings for other bulk physics measurements.
The rest of this paper discusses femtoscopic measurements, elliptic flow and strangeness yields and uses
entropy scalings to make naive predictions for these observables at the LHC.

\section{Femtoscopy}

\begin{figure}[htbp]
    \begin{center}
  \includegraphics[width=\linewidth]{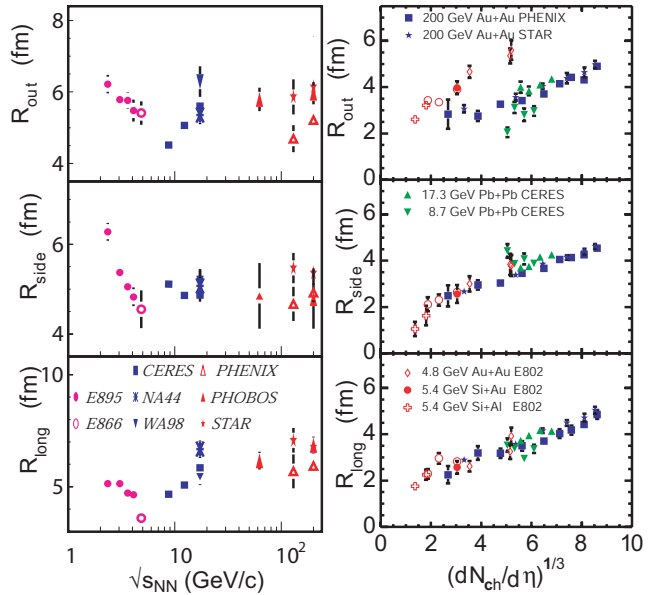}\\
  \caption{Left panel) World dataset of published HBT radii from central Au+Au (Pb+Pb) collisions versus collision energy.
  Compilation from~\cite{HBTComp}. Right panel) The same measurements as in the left panel but as a
  function of (dN$_{ch}$/d${\eta})^{1/3}$}\label{Fig:HBT}
  \end{center}
\end{figure}

Femtoscopic measurements examine space-momentum correlations and allow, in
a model dependent fashion, for the extraction of source sizes. For more details on these techniques see elsewhere
\cite{HBT}. Detailed measurements have been made, especially looking into identical pion correlations. While the
radii hover around 5 fm,  when the results are plotted as a function of \sqrts, see Fig.~\ref{Fig:HBT} left panel~\cite{LHCdndy},
 no clear trends appear. However, if, as in Fig.~\ref{Fig:HBT} right panel, the same measurements are now plotted
 as a function of (dN$_{ch}$/d${\eta})^{1/3}$ a clear scaling is observed for R$_{long}$ and R$_{side}$. The applicability
 to R$_{out}$ is less clear. This lack of scaling for R$_{out}$ might have an explanation via the fact
 that this HBT variable consists of a convolution of terms containing length and time scales. R$_{long}$ and R$_{side}$
are, on the other hand, pure length measurements. The deviations in R$_{out}$ might therefore be signatures of differing timescales for the
collisions. It should be noted that while the right panel of Fig.~\ref{Fig:HBT} is for  $<k_{T}>$ = 400 MeV and 390 MeV for
RHIC and SPS respectively. As good a scaling is observed if other  $<k_{T}>$ ranges are selected. The non-monotonic
behavior seen in the left panel of Fig.~\ref{Fig:HBT}  for \sqrts $<$ 5 GeV makes it obvious that
 the dN$_{ch}$/d$\eta$ scaling  breaks down at very low collision energies. It has been pointed out, by the CERES collaboration,
  that this is probably due to the dominance of baryons at lower \sqrts~\cite{HBTCeres}.

\section{Elliptic Flow}

\begin{figure}[htbp]
    \begin{center}
  \includegraphics[width=\linewidth]{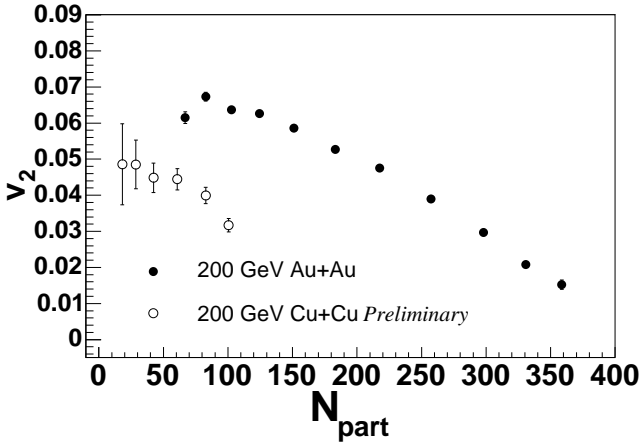}\\
  \caption{Elliptic flow, v$_{2}$, measured near mid-rapidity as a function of N$_{part}$ for Cu-Cu (open symbols)
  and Au-Au (closed symbols) collisions at \sqrts= 200 GeV. Only statistical errors are shown, taken from \cite{Scaledv2}. }\label{Fig:CuCuv2}
  \end{center}
\end{figure}

Peripheral and low energy collisions are likely to produce systems with incomplete thermalization.
Since  re-scattering of the particles is rare, in this low density limit regime,  little
change occurs, on average, to the initial momentum distributions. The measured elliptic flow, v$_{2}$, is therefore
proportional to the initial state eccentricity, $\epsilon$, as determined via Glauber calculations, and the
space density of the initial particles. The latter determining the scattering probability which  must be
finite so that the  initial spatial anisotropy can be converted into a measurable momentum anisotropy. Thus,

\begin{eqnarray}
v_{2} \propto \frac{1}{Area}\frac{dN}{dy} \epsilon \\
\epsilon =  \frac{(R_{x}^{2}-R_{y}^{2})}{(R_{x}^{3}+R_{y}^{2})}\label{LowDenLimit}
\end{eqnarray}

where R$_{x}$ and R$_{y}$ are the major and minor axes of the overlap ellipse of the colliding nuclei in
the transverse plane. From Eqn.~\ref{LowDenLimit} it can be seen that in the most central A-A collisions
$\epsilon$, and hence v$_{2}$,  tends towards zero. In the Cu-Cu
data however there remains a significant measured v$_{2}$ component in the most central events,
Fig.~\ref{Fig:CuCuv2}, \cite{Scaledv2}.

~

\begin{figure}[htbp]
 \begin{center}
  \includegraphics[width=\linewidth]{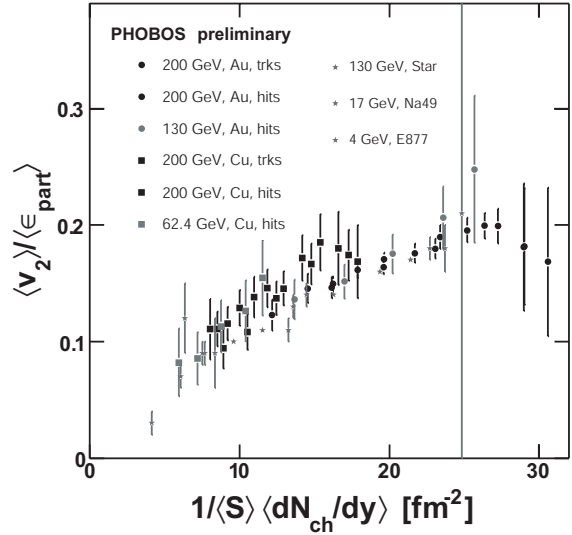}\\
  \caption{v$_{2}$ per average participant eccentricity as a function of mid-rapidity area density. Data are
  from RHIC  Au-Au collisions at \sqrts=130 and 200 GeV and Cu-Cu collisions at \sqrts = 62.4 and 200 GeV as well
  as SPS \sqrts= 17 GeV and AGS \sqrts = 4 GeV collisions~\cite{Scaledv2}. }\label{Fig:Elliptic}
   \end{center}
\end{figure}

Also when, for a given number of participants, N$_{part}$,  v$_{2}$/$\epsilon$ measurements in  Cu-Cu are
 contrasted to similarly scaled Au-Au  results,
 they are not compatible. This suggests that the low density limit
concept is not applicable to the data~\cite{CuCuv2}.
  Since the Cu-Cu results are above those of the Au-Au at a given N$_{part}$, other conclusions
   could be drawn from these data. These are that
 either the Cu-Cu system is more efficient at translating spatial anisotropies into v$_{2}$ signals
or there are significant non-flow effects in the Cu-Cu data that create these large momentum asymmetries.
An alternative explanation can however be postulated by examining the initial state eccentricity calculation. As stated above
this is traditionally defined as an average eccentricity calculated from the participant nucleon distributions
relative to the reaction plane for each centrality class via Glauber calculations. This method leads to two
possible errors:  firstly event-by-event fluctuations are  averaged to zero  and
secondly, while the  minor axis  of the participant distribution  is along the impact parameter  vector on average
 for any given event this will not generally be the case. In events with a small number of participants these
 two effects can cause a significant error in the eccentricity calculations~\cite{Scaledv2}. The PHOBOS collaboration
have therefore proposed a new technique for calculating the eccentricity, called $\epsilon_{part}$, which takes these
effects into account \cite{CuCuv2}. When the v$_{2}$/$\epsilon_{part}$ ratio is then plotted versus the mid-rapidity
area density of produced particles, Fig.~\ref{Fig:Elliptic}, good agreement is now observed between the two collision
 systems. Also shown in this plot are results from \sqrts = 17 and 4 GeV and a common scaling seems to emerge. This
 suggests that v$_{2}$ depends solely on the initial area density achieved in the collision over two orders of
 magnitude in collision energy. Note that a linear
 dependence on dN/dy or entropy has emerged.

\section{Strange particle production}

A lack of phase space in \pp collisions is predicted to result
in strangeness production being suppressed~\cite{Redlich}. The magnitude
of this suppression decreases with increasing \sqrts and with the volume of
the medium produced. Since the collision volume is linearly proportional to N$_{part}$,
it is also predicted that the strangeness production volume follows such a linear
scaling. The calculations also show that the \pp suppression increases with
the particle's strangeness content. When the system becomes sufficiently large, the
phase space restrictions are removed and strangeness can be produced freely.
For these types of collisions, the ratio of strangeness yield to N$_{part}$
is therefore a constant as a function of centrality.

\begin{figure}[htbp]
    \begin{center}
  \includegraphics[width=\linewidth]{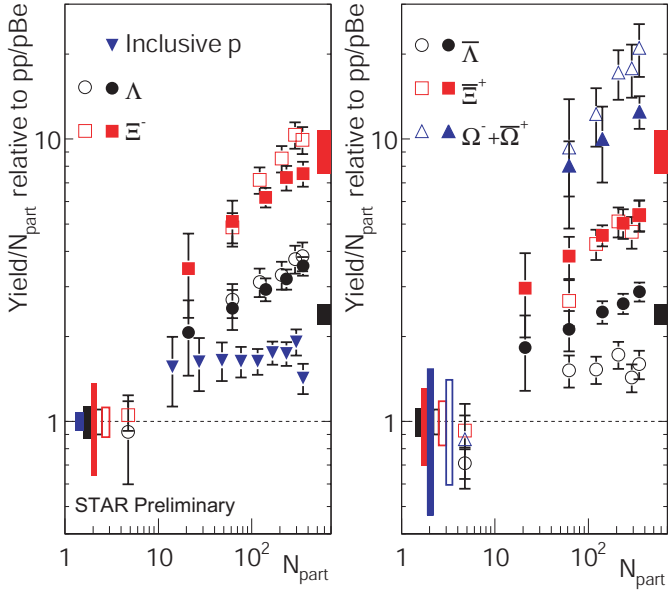}\\
  \caption{ Yield per participant relative to the yield per participant
  in \pp collisions  as a function of N$_{part}$. Solid symbols are for Au-Au collisions at \sqrts = 200 GeV,  and the
open symbols are measured by NA57 from Pb-Pb at \sqrts=17.3 GeV. The
boxes represent the combined statistical and systematical
uncertainties in the \pp and p-Be data. The error bars on the
data points represent those from the heavy-ion measurement. The boxes on the
right axes mark the predictions from a model using a Grand Canonical
formalism described in \cite{Redlich}. }\label{Fig:StrangeNpart}
  \end{center}
\end{figure}

It can  been seen in Fig.~\ref{Fig:StrangeNpart} that
in Au-Au collisions at \sqrts = 200 GeV, although the predicted strangeness ordering
of the \pp suppression is observed the yields per participant are not
constant. Further, the magnitude of the suppression for a given strange baryon
is the same in \sqrts = 200 GeV collisions as at the SPS in \sqrts= 17.3 GeV collisions~\cite{EnhanceStar,EnhanceNA57}.
Both of these results are counter to
the predictions. Calculations using statistical models applied to the STAR
data indicate that $\gamma_{s}$, the strangeness saturation factor, is unity in
central collisions, e.g. that there is no suppression in the most central
data~\cite{StarSpectra}. This suggests that the
relevant volume for strangeness production is not linearly
proportional to N$_{part}$ and hence not purely related to  the
initial collision overlap volume. Therefore  some other scaling must be sought.

\begin{figure}[htbp]
    \begin{center}
  \includegraphics[width=\linewidth]{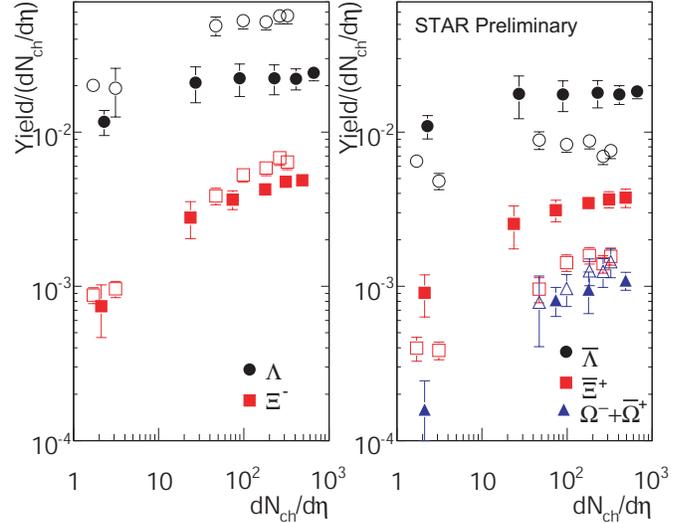}\\
  \caption{ Yield per dN$_{ch}$/d${\eta}$ as a function of dN$_{ch}$/d${\eta}$ for strange
(anti)baryons. Solid symbols are for Au-Au collisions at \sqrts = 200 GeV,  and the
open symbols are measured by NA57 from Pb-Pb at \sqrts=17.3 GeV~\cite{EnhanceNA57}. }\label{Fig:StrangeNch}
  \end{center}
\end{figure}

~

Figure \ref{Fig:StrangeNch}  shows the yield per dN$_{ch}$/d${\eta}$ for $\Lambda$, $\bar{\Lambda}$,
$\Xi^-$, $\bar{\Xi}^{+}$, and $\Omega^{-}$+$\bar{\Omega}^{+}$ for RHIC and the SPS Pb-Pb collisions using data
taken from \cite{EnhanceStar,EnhanceNA57}. When plotted in this way the yield/dN$_{ch}$/d${\eta}$
appears constant for the more central data suggesting that an entropy measure is more closely
correlated to the strangeness production volume than the number of participants.

\section{LHC predictions}

\begin{figure}[htbp]
 \begin{center}
  \includegraphics[width=\linewidth]{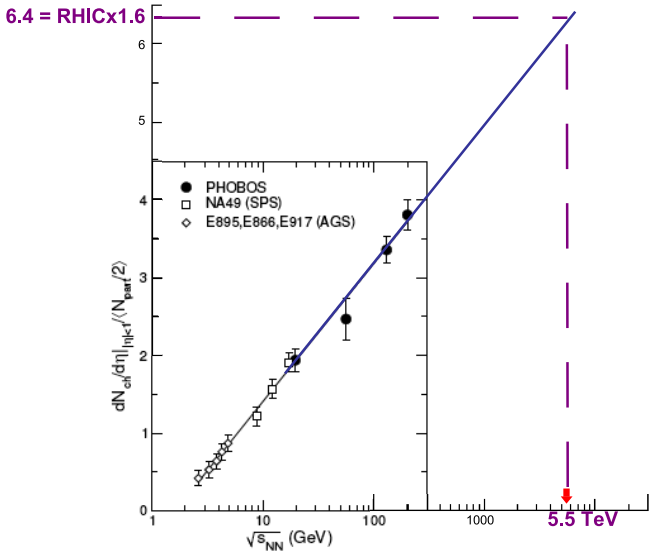}\\
  \caption{ The multiplicity density per participant pair measured in Au-Au and Pb-Pb collisions at AGS, SPS, and RHIC.
  Taken from~\cite{PionSqrts}, and naively extrapolated to  LHC energies in \cite{LHCdndy}.}\label{Fig:LHCdNdy}
   \end{center}
\end{figure}

As shown in Fig.~\ref{Fig:Entropy}, it appears that the EoS for heavy-ion collisions  is different to that
in \pp. We also expect it to be much more complicated than that derived earlier for a gas of massless pions.
There can still be a functional relationship between   entropy and \sqrts. In Fig.~\ref{Fig:LHCdNdy},
 taken from \cite{PionSqrts},  measurements of the mid-rapidity dN$_{ch}$/d${\eta}$ per participant pair are shown
as a function of collision energy, note the log scale on the abscissa. This plot clearly shows a linear scaling
of N$_{ch}$/d${\eta}$ per participant pair with log(\sqrts) again over several orders of magnitude. This dramatic scaling
allows an estimation of the change in entropy as a function of beam energy to be made.

  Making a leap of faith, one can  naively extend this scaling to the LHC energy regime of 5.5 TeV, as performed in
   \cite{LHCdndy}, and  shown by the solid line in Fig.~\ref{Fig:LHCdNdy}. Such an extrapolation
leads to the expectation that mid-rapidity multiplicities  will be as little as 60$\%$ higher than they are at RHIC,
e.g. the mid-rapidity yield will be of order 1200 charged particles.  Armed with this ``knowledge" of the expected
LHC multiplicity/entropy one can continue to make simple extrapolations. I use the scalings with entropy
 for the other bulk physics measurements discussed above. Thus predictions for the HBT radii, elliptic flow
 and strangeness yields are obtained.

~

 The femtoscopic radii  were shown to depend primarily on the cube root of the mid-rapidity event multiplicity.
Thus the combination of  Figures~\ref{Fig:HBT} and~\ref{Fig:LHCdNdy} suggests that HBT measurements
in central collisions at LHC will be  $\sim 17\%$ higher than at RHIC ($(1.6)^{1/3} = 1.17$).
Precise expectations for $R_{out}$   at the LHC  are, admittedly, less certain, as the former does not
scale exactly with multiplicity (Fig.~\ref{Fig:HBT}).

~

To make a prediction for the value of elliptic flow in the most central collisions one needs a calculation of the
mid-rapidity transverse area. For Pb-Pb collisions Glauber calculations show an initial RMS transverse radius of
$\sim$2.8 fm and hence a mean particle area density of $\sim$ 50 fm$^{-2}$. Using a straight line extrapolation from
the data shown in Fig.~\ref{Fig:Elliptic}, one would predict v$_{2}/\epsilon_{part}$ = 0.35. However, it should be
remembered that the proportionality of elliptic flow to the mean transverse area density is only predicted to hold
in the low particle density limit. It is hard to believe that central Au-Au or Pb-Pb collisions produce low area
densities and there is much evidence that the re-scattering rate is high at RHIC, for instance the large amount of radial
flow~\cite{RadialFlowStar,RadialFlowPhenix}
 and the significant regeneration of resonances after chemical freeze-out~\cite{Resonances}. Taking these, and other,
  RHIC results into consideration it is more likely that central RHIC/LHC collisions are hydrodynamical
  in behavior. If this is the case, v$_{2}$/$\epsilon$ will be a constant, assuming the same EoS holds at
  RHIC and the LHC. If one takes a second look at
  Fig.~\ref{Fig:Elliptic} it is possible that the higher centrality Au-Au data points are pulling away from a
  straight line fit towards a plateau. Therefore instead of LHC collisions increasing further the elliptic flow to
  eccentricity ratio  is likely to stay constant at v$_{2}$/$\epsilon$  = 0.2.

  ~

Although the strangeness yields cannot be  directly extracted from Fig.~\ref{Fig:StrangeNch}, an estimate can still
be made for the LHC regime. From Fig.~\ref{Fig:StrangeNch} it can be seen that whereas the $\bar{\Lambda}$/$\Lambda$ ratio
at the SPS  is $\sim$0.1, at RHIC the mid-rapidity anti-baryon to baryon
ratio  is very close to unity. It can
also be seen that the SPS strangeness yields per N$_{ch}$/d${\eta}$ for the (anti)$\Lambda$ and (anti)$\Xi$ frame
those measured at RHIC. These results suggest therefore that the LHC net-baryon number will be essentially zero
and that the yield per charged particle at mid-rapidity will be the same as that measured at \sqrts = 200 GeV.
Given that the entropy extrapolations predict an $\sim$60$\%$ increase in charged particle production one can therefore
expect a similar increase in the  strange baryon yields when going from RHIC to the LHC. Thus,
dN$_{\Lambda}$/dy = dN$_{\bar{\Lambda}}$/dy = $\sim 20-30$, dN$_{\Xi^{-}}$/dy = dN$_{\bar{\Xi}^{+}}$/dy = $\sim 4-6$ and
dN$_{\Omega^{-} +\bar{\Omega}^+}$/dy = $\sim$ 0.5-1 in the most central events.

\section{Summary}

In summary, from simple arguments it can be shown that produced particle multiplicities are strongly
correlated to the entropy of the system and that a limiting temperature is reached if the  matter being probed
is  hadronic. Further, when heavy-ions are collided at high  energies, the trend of this
entropy measure versus the Fermi energy is different to that in \pp  and low \sqrts heavy-ion collisions.

~

Evidence is mounting that many observables from the  soft-sector
  are determined primarily by total multiplicity, i.e. entropy.
 The  HBT radii, properly-scaled elliptic flow,  and strangeness yields, all appear to show universal multiplicity scalings.
This  implies that bulk observables are dominated by  entropy-driven factors.
There is also much evidence that these high-energy collisions produce  a new phase of matter that consists of
interacting quarks and gluons. That this highly non-trivial source apparently  produces signatures, in the soft momentum sector,
with  dependencies only on multiplicity and not reaction energy is very intriguing.

~

These multiplicity scalings can then be used to make naive predictions for some soft sector results at the LHC.
However, anything can, of course occur when this new
 regime is probed as novel physics could be probed. However, any significant deviation from these apparent universal
multiplicity scalings  would be extremely. interesting. Whatever the outcome the first results from CERN, now only
just over a year away, are eagerly awaited.

\end{document}